\RequirePackage{fix-cm}
\documentclass[twocolumn,final]{svjour3}        % twocolumn
\usepackage{filecontents}
\smartqed  % flush right qed marks, e.g. at end of proof
\usepackage{graphicx}
 \usepackage{mathptmx}      % use Times fonts if available on your TeX system
\usepackage{color}
\usepackage{latexsym}
\usepackage{xspace}
\usepackage[hyphens]{url}
\usepackage[hidelinks]{hyperref}
\hypersetup{breaklinks=true}
\usepackage{cite}

\def\lesssim{\mathrel{\hbox{\rlap{\hbox{\lower4pt\hbox{$\sim$}}}\hbox{$<$}}}}
\def\gtrsim{\mathrel{\hbox{\rlap{\hbox{\lower4pt\hbox{$\sim$}}}\hbox{$>$}}}}
\def\alt{\mathrel{\hbox{\rlap{\hbox{\lower4pt\hbox{$\sim$}}}\hbox{$<$}}}}
\def\agt{\mathrel{\hbox{\rlap{\hbox{\lower4pt\hbox{$\sim$}}}\hbox{$>$}}}}

\definecolor{darkgreen}{rgb}{0,0.6,0}
\newcommand{\co}[1]{\textcolor{black}{#1 }}

\newcommand{\newacronym}[3]{%
  \newcommand{#1}{#2 (#3)\xspace%
    \renewcommand{#1}{#3\xspace}%
  }%
}
\usepackage[normalem]{ulem}

\newcommand{\gstlal}{\texttt{gstLAL}\xspace}
\newcommand{\htcondor}{HTCondor\xspace}

\newcommand{\glidein}{Glidein\xspace}

\newacronym{\ldg}{LIGO Data Grid}{LDG}
\newacronym{\cvm}{CERN Virtual Memory File System}{\texttt{CVMFS}}
\begin{document}

\title{Supporting High-Performance and High-Throughput Computing for Experimental Science}

%\thanks{Grants or other notes
%about the article that should go on the front page should be
%placed here. General acknowledgments should be placed at the end of the article.}
%\subtitle{Do you have a subtitle?\\ If so, write it here}

%\titlerunning{Short form of title}        % if too long for running head

\author{E.~A.~Huerta \and Roland~Haas \and Shantenu~Jha  \and Mark~Neubauer \and Daniel~S.\ Katz}

%\authorrunning{E.\ A.\ Huerta \and Roland~Haas \and Shantenu~Jha \and Mark~Neubauer \and Daniel~S.\ Katz} % if too long for running head

\institute{E.\ A.\ Huerta  \at
              National Center for Supercomputing Applications \& \\
              Department of Astronomy, University of Illinois at Urbana-Champaign, 
              Urbana, Illinois 61801, USA\\
              \email{elihu@illinois.edu}           %  \\
%             \emph{Present address:} of F. Author  %  if needed
           \and
           Roland Haas \at
           National Center for Supercomputing Applications, University of Illinois at Urbana-Champaign, 
              Urbana, Illinois 61801, USA\\
              \email{rhaas@illinois.edu} 
              \and
           Shantenu~Jha \at
           Brookhaven National Laboratory and Rutgers,  The State University of New Jersey, Piscataway, New Jersey, 08854\\
              \email{shantenu.jha@rutgers.edu}
           \and
           Mark~Neubauer \at
           Department of Physics \& National Center for Supercomputing Applications, University of Illinois at Urbana-Champaign, 
              Urbana, Illinois 61801, USA\\
              \email{msn@illinois.edu}    
          \and
           Daniel~S.\ Katz \at
           National Center for Supercomputing Applications \& Department of Computer Science \& Department of Electrical and Computer Engineering \& School of Information Sciences, University of Illinois at Urbana-Champaign, 
              Urbana, Illinois 61801, USA\\
              \email{dskatz@illinois.edu}     
}

\date{Received: 6 October 2018 / Accepted: 24 January 2019}
% The correct dates will be entered by the editor

\maketitle
\begin{abstract}

The advent of experimental science facilities---instruments and observatories, such as the Large Hadron Collider (LHC), the Laser Interferometer Gravitational Wave Observatory (LIGO), and the upcoming Large Synoptic Survey Telescope (LSST)---has brought about challenging, large-scale computational and data processing requirements.
Traditionally, the computing infrastructures to support these facility's requirements were organized into separate infrastructure that supported their high-throughput needs and those that supported their
high-performance computing needs. We argue that in order to enable and accelerate scientific discovery at the scale and sophistication that is now needed, this separation 
between High-Performance Computing (HPC) and High-Throughput Computing (HTC) 
must be bridged and an integrated, unified 
infrastructure must be provided. In this paper, we discuss several case studies where such infrastructures
have been implemented. These case studies span different science domains, software systems, and application
requirements as well as levels of \co{sustainability.} A further aim of this paper is to provide a basis to determine the common characteristics and requirements of such infrastructures, as well as to begin a discussion 
of how best to support the computing requirements of existing and future experimental science
facilities.

%Insert your abstract here. Include keywords, PACS and mathematical
%subject classification numbers as needed.
\keywords{HPC \and HTC \and LIGO \and CMS \and ATLAS \and Blue Waters \and Titan \and \texttt{OSG} \and containers}
% \PACS{PACS code1 \and PACS code2 \and more}
% \subclass{MSC code1 \and MSC code2 \and more}
\end{abstract}

\section{Introduction}
\label{intro}

To discuss high performance computing (HPC) and high throughput computing (HTC), we initially need to distinguish between ``computing modes'' and ``computing infrastructure'', as the terms HTC and HPC are often used for both. HTC as a computing mode is typically
used for workloads that are primarily characterized by the number of tasks associated
with the workload. HTC workloads are comprised of tasks that are typically independent of each other, that is to say, the tasks can start or complete in any order.
Furthermore, while a task (defined as a unit of work) is most often equated to a job (defined as an entity submitted to a regular batch queue), a single task does not need to be mapped to a job; multiple tasks might also be mapped into a single job. In contrast, an HPC workload is characterized by a metric such as its scalability or some other measure of performance (e.g., number of flops). Typically an HPC
workload comprises a single task that is executed as single job; however, 
HPC workloads might comprise multiple tasks with dependencies but may still be packed as a single job.
\co{Similarly each HTC tasks typically operates on a small volume of data
only, even though the total amount of data processed by the workload is often
quite large, as in big-data applications. Often data needs to be staged from
external sources before computation can begin, and it is the infrastructure's
responsibility to ensure that all required input data is present at the time
computing begins, while during the runtime of a single task, IO requirements
are typically small. A prototypical HPC workload on the other hand will
operate on a single, large volume of data, typically in the form of initial
data files or checkpoint files which are provided by the user or written by a
previous task in the HPC workload. Since IO is typically a synchronization
point in an HPC workload, insufficient IO bandwidth in even a subset of the
compute resources used by the workload becomes a significant bottleneck for
the workload as a whole.}
These distinctions are important to appreciate the different workloads that are covered in this paper, and as the astute reader will notice, some workloads defy reduction into one or the other category.

Furthermore, there are computing infrastructures that are designed
to primarily support one type of workload. The canonical example is Condor-based~\cite{condor-practice}
systems whose design point is to maximize the number of tasks per unit time, known as throughput.
On the other hand, most supercomputers and high-performance clusters that have high-performance interconnects and memory are typically designed for HPC workloads. Traditionally, HTC workloads have not been executed on such HPC infrastructures, but as this paper illustrates, there have been many recent attempts to run
HTC workloads on HPC infrastructures. \co{The motivation for this work is multifold, and has its roots at the creation of the Open Science Grid (\texttt{OSG})~\cite{pordes2007open}, a network of computing centers designed to share resources and data automatically, that was created to meet the ever increasing computing needs of US researchers utilizing the Large Hadron Collider (LHC) at CERN. At present, core members of the OSG project include the ATLAS and CMS Collaborations, and long-time stakeholders such as the Laser Interferometer Gravitational wave Observatory (LIGO)~\cite{DII:2016,LSC:2015} Scientific Collaboration.}

\co{This paper chronicles how these disparate communities have developed software systems that 
that by virtue of being compatible with the \texttt{OSG}, thereby requiring the use of containerized software stacks, can be seamlessly ported into HPC platforms. Furthermore, since workloads compatible with \texttt{OSG} good practices and policies consist of a large number of embarrassingly parallel jobs, they present an ideal case to boost utilization of HPC compute nodes that would otherwise remain idle. This approach is not only strongly encouraged but also cost-effective for two reasons: (i) HTC-type workloads increase cluster utilization and throughput of HPC platforms by backfilling nodes that would otherwise remain idle; and (ii) interoperability of HTC and HPC platforms enables an optimal use of existing cyberinfrastructure facilities, and provides scenarios that may inform the design and construction of future supercomputing facilities. This paradigm has benefited the scientific and HPC/HTC communities, and has played a central role in pushing the boundaries of our knowledge in high energy physics and gravitational wave astronomy, leading to remarkable discoveries that have been recognized with the Nobel Prizes in Physics in 2013 and 2017.} 

This article is organized around a set of case studies. Each study, in Section~\ref{sec:studies}, briefly describes the science problem, and the rationale to go beyond available HTC infrastructures. Section~\ref{htc_on_hpc} describes how HPC has been incorporated at both the middleware/system and application levels, and what the impact of the work will be on the science. In Section~\ref{sec:analysis} we compare and briefly analyze the different approaches. In Section~\ref{sec:5} we discuss the future of this program in the context of exascale computing and emergent trends in large scale computing and data analytics that leverage advances in machine and deep learning.

\section{Case Studies}
\label{sec:studies}

This section presents a brief overview of two large-scale projects, gravitational wave astrophysics and high energy physics, that initially met their data analysis needs with workloads that were tailored to use HTC platforms. We discuss how the computational needs of these science missions led to the construction of a unified HTC-HPC infrastructure, and the role of \texttt{OSG} and containers to facilitate and streamline this process. 

\subsection{Gravitational wave astrophysics: from theoretical insights to scientific discovery}
\label{sec:ligo}

According to Einstein's theory of general relativity, gravity is a manifestation of
spacetime curvature~\cite{gr}. Gravitational waves are generated when masses are accelerated 
to velocities closer to the speed of light. Gravitational waves remove 
energy from the system of masses, which translates into a rapid shrinkage of 
the orbital separation between the masses, and culminates in a cataclysmic collision
accompanied by a burst of gravitational radiation~\cite{linear,quadrupole}. 

Over the last two years, LIGO~\cite{DII:2016,LSC:2015} and its European parter Virgo~\cite{Virgo:2015}, have made
ten gravitational wave detections that are consistent with the merger of two
black holes~\cite{DI:2016,secondBBH:2016,thirddetection,fourth:2017,GW170608,o1o2catalog}. The trail of discovery has also led to the first direct detection of two colliding neutron stars~\cite{bnsdet:2017}, which was observed with two cosmic messengers: gravitational waves and light. This multimessenger observation has provided evidence that the collision of neutron stars are
the central engines that trigger short gamma ray bursts, the most energetic electromagnetic explosions
in the Universe after the Big Bang, and the cosmic factories that where about half 
of all elements heavier than iron are produced~\cite{bnsdet:2017,mma:2017arXiv,2017arXiv171005836T}.

To understand the physics of gravitational wave sources and enable their discovery, a worldwide, decades-long research program was
pursued to develop 
numerical methods to 
solve Einstein's general relativity equations in realistic astrophysical 
settings~\cite{smarr,1989fnr,1993PhRvLA,1995Sci270941M}. The
computational expense and scale of these numerical relativity simulations 
require large amounts of
computing power. The lack of such computing power to address this physics
problem is one of the elements that led to the foundation of the US National Science Foundation (NSF) supercomputer centers, including the National Center for Supercomputing Applications (NCSA).

\begin{figure}[htbp]
\centering
%\includegraphics[width=0.48\linewidth]{horizon_and_waves_grid_2k}
%\hfill
\includegraphics[width=0.48\linewidth]{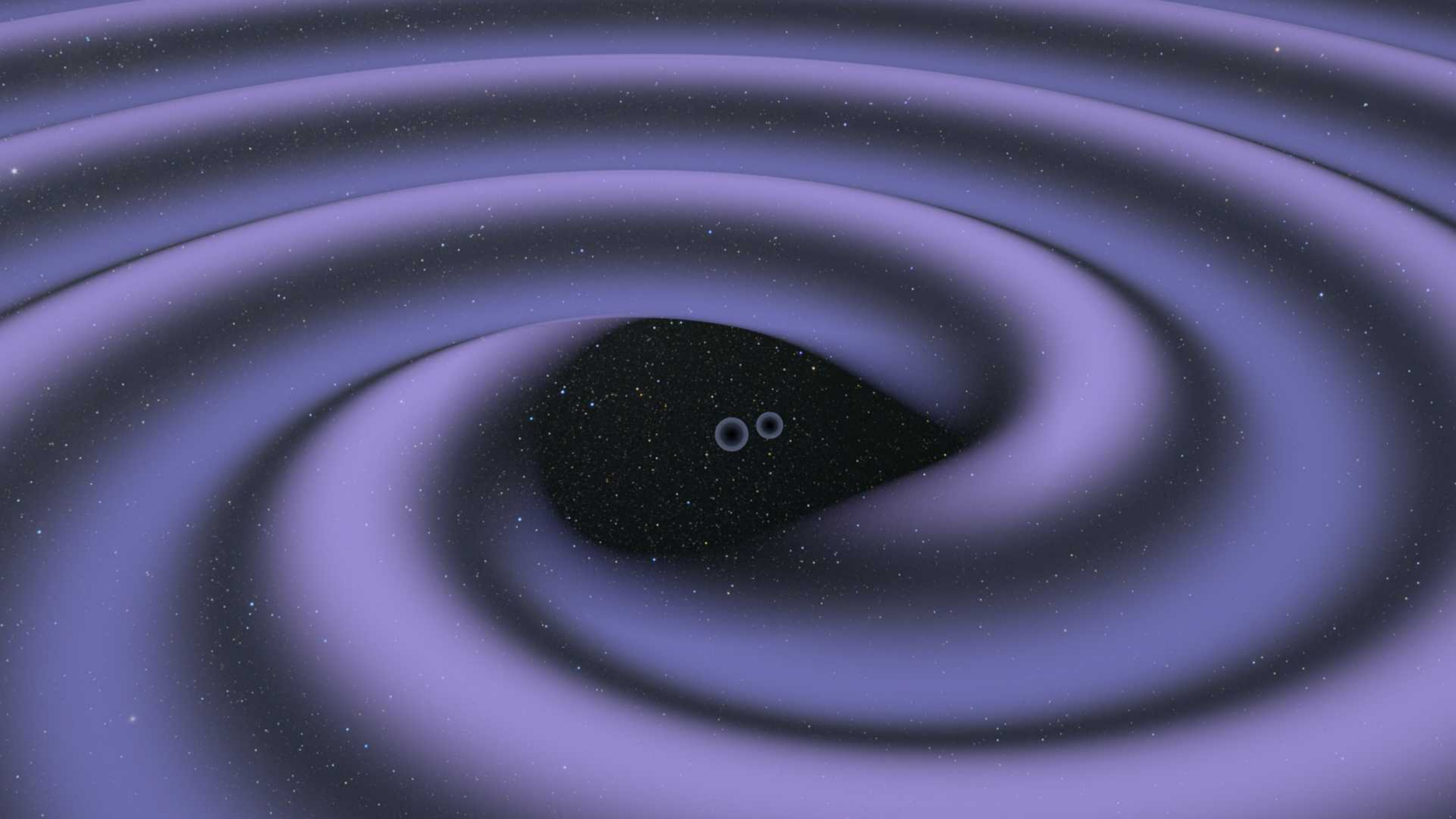}
\includegraphics[width=0.48\linewidth]{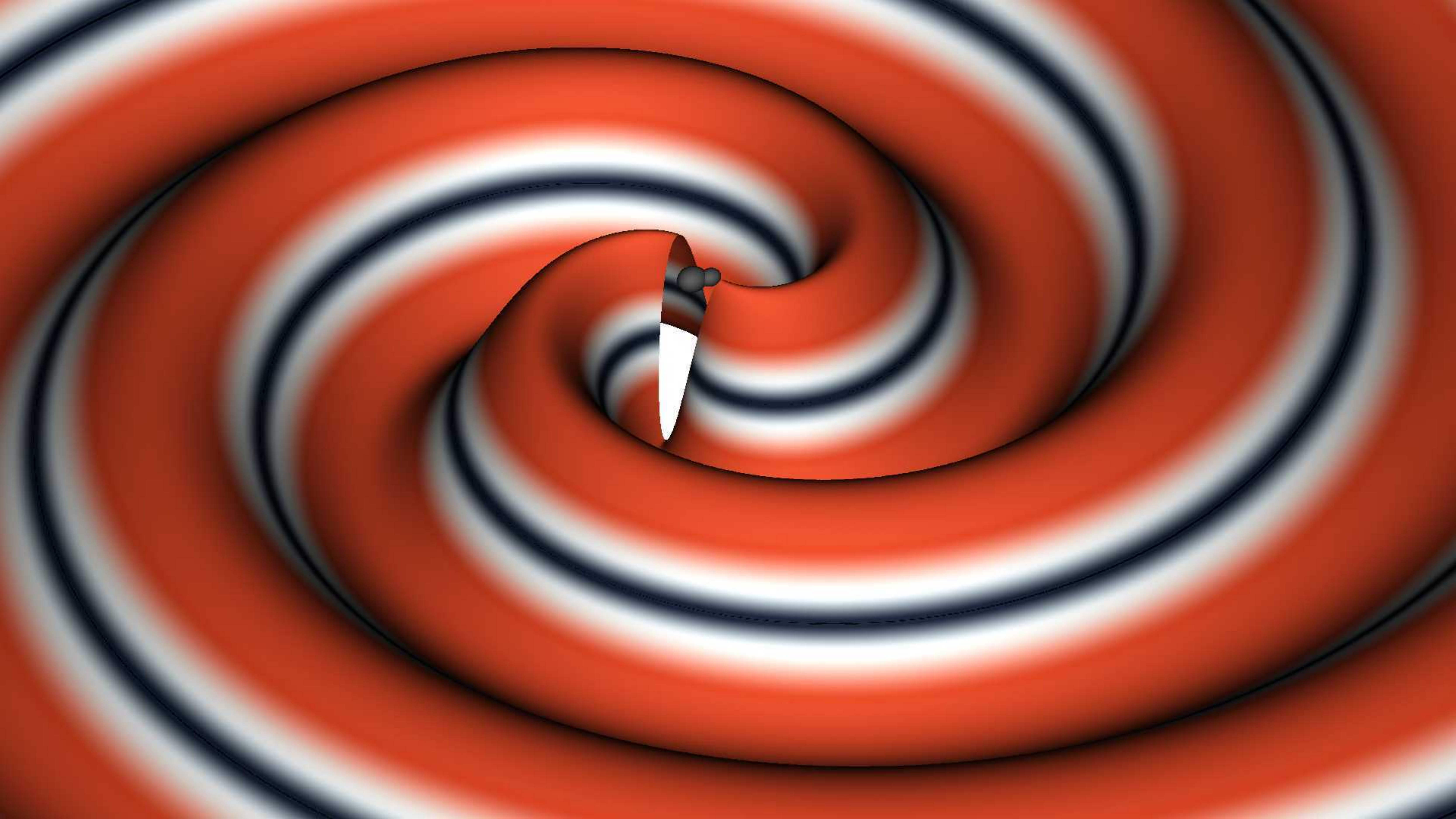}
\caption{Visualization of the event horizons and gravitational waves emitted
by the first~\cite{DI:2016} and fourth~\cite{fourth:2017} pair of merging black holes detected by LIGO. These gravitational waves induce changes in 
the arm length of the LIGO and Virgo detectors that are smaller than the diameter of a proton.
}\label{fig:bbh-vis}
\end{figure}

The first numerical evolutions of two orbiting black holes 
that inspiral into each other and
eventually merge were reported in 2005~\cite{preto}. This breakthrough was 
reproduced independently by other groups shortly thereafter using entirely different 
software stacks~\cite{baker:2006,camp:2006}. 
From that point onwards, numerical relativists embarked on a vigorous program to
produce mature software that could be used to routinely simulate the merger of black holes
in astrophysically motivated settings. Figure~\ref{fig:bbh-vis} shows still images
of black hole collisions that represent a sample of the black hole mergers detected by
the LIGO detectors, which we numerically simulated using the open source, 
Einstein Toolkit~\cite{naka:1987,shiba:1995,baum:1998,baker:2006,camp:2006,Lama:2011,wardell_barry_2016_155394,ETL:2012CQGra,Ansorg:2004ds,Diener:2005tn,Schnetter:2003rb,Thornburg:2003sf} community software on the Blue Waters supercomputer~\cite{bluewaters:web,Kramer2015,2017arXiv170300924J}. 

\begin{figure}[htbp]
\centering
\includegraphics[width=0.8\linewidth]{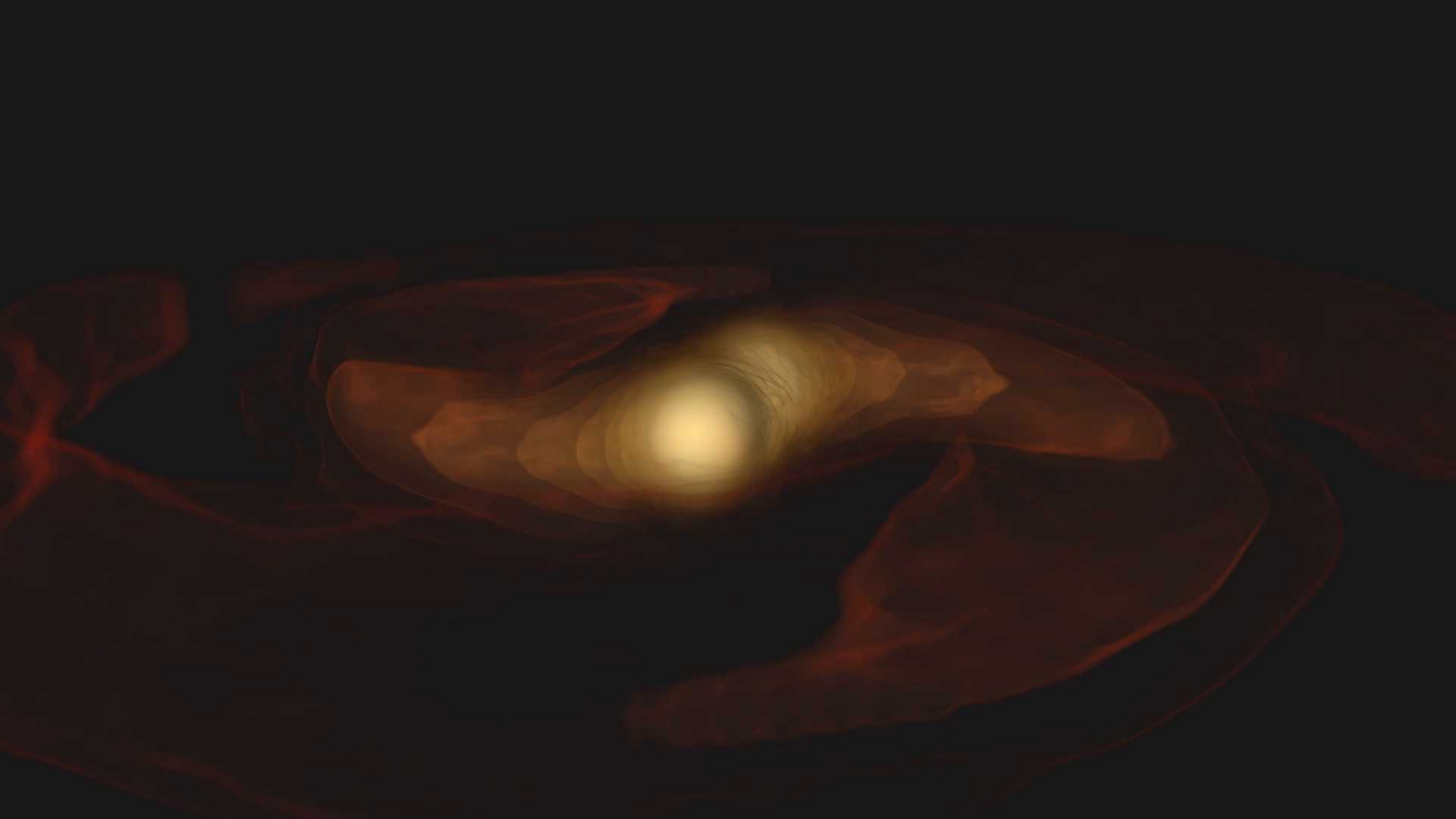}
\caption{Visualization of the merger of two neutron stars.
This simulation is consistent with the astrophysical properties of the two
colliding neutron stars detected by the LIGO and Virgo detectors. 
}\label{fig:bns}
\end{figure}

\noindent While the numerical modeling of black hole collisions has evolved rapidly over the last decade,
the modeling of astrophysical objects that involve matter, such as neutron star
collisions, has progressed at a lower pace~\cite{2015CQGra..32q5009E,2016PhRvD..93l4062H,2014CQGra..31a5005M,ETL:2012CQGra,2017JCoPh.335...84K}. The different timescales involved in these
complex systems, and the need to couple Einstein's field equations with magneto--hydrodynamics and 
microphysics is a challenging endeavor. Recent efforts to cross validate the physics described
by different software stacks is an important step towards the development of mature software 
that can be routinely used to simulate these events. This research program is timely and relevant given 
that LIGO, Virgo, and several astronomical facilities are coordinating efforts to identify new
multimessenger events in the upcoming LIGO-Virgo gravitational wave discovery campaign, known
as O3. Figure~\ref{fig:bns} shows one of the numerical relativity simulations we produced
to numerically model the neutron star collision detected by the LIGO and Virgo detectors. The simulation
was produced with the GRHydro numerical relativity code~\cite{2014CQGra..31a5005M} on the Blue Waters supercomputer.

\noindent Available catalogs of numerical relativity simulations~\cite{Mroue:2013} have been used to calibrate semi-analytical 
waveform models that are utilized during gravitational wave discovery campaigns~\cite{Bohe:2016gbl,husacv:2016PhRvD,khan:2016PhRvD,Tara:2014}. This is because generating
numerical relativity waveforms takes between several days (black hole mergers) and several months 
(neutron star mergers) in HPC infrastructures. However, since gravitational wave detection requires low latency
analyses of gravitational wave data, numerical relativity catalogs are used to calibrate models that can 
generate simulated waveform signals in tens of milliseconds. 

Once a new gravitational wave trigger is identified,  numerical relativity catalogs that actually reproduce the signals extracted from
LIGO and Virgo data are created. To inform this analysis, and constrain the region of interest in the
8-dimensional parameter space that describes gravitational wave sources, 
LIGO and Virgo data is carefully analyzed using robust Bayesian algorithms~\cite{bambiann:2015PhRvD}. 
With these catalogs of numerical relativity waveforms, it is possible
to infer the astrophysical origin and environments of gravitational wave sources.

In conclusion, the numerical modeling of gravitational wave sources, and the validation 
of new discoveries with numerical relativity waveforms,
depends critically on HPC infrastructure.

\subsubsection*{Gravitational Wave Detection with HTC workloads on HTC infrastructures}

The choice of signal-processing techniques for
gravitational wave detection has produced
workloads that
are computationally expensive and poorly scalable. 
These algorithms sift through gravitational wave data, 
looking for a high correlation with modeled waveform templates,
which are calibrated with numerical relativity waveforms. If this template
matching method finds a noise trigger with high significance, then
it is followed up using a plethora of statistical algorithms to ensure
that it is not a noise anomaly, but an actual gravitational wave
candidate that is observed in several gravitational wave detectors. 

In a typical discovery campaign, LIGO utilizes template banks that have
$\sim$$10^5$ distinct modeled waveforms. Each segment of gravitational wave data,
$400\,\mathrm{MB}$ in size, is matched-filtered against 
every single one of these template waveforms. 

LIGO employs two separate pipelines, 
\texttt{PyCBC}~\cite{2016CQGra..33u5004U} and \gstlal~\cite{2012ApJ...748..136C}, to perform
matched-filtering based gravitational waves. Both pipelines use
\htcondor~\cite{condor-practice} as a
workflow management system. To provide computing power for these computationally intensive 
searches, LIGO maintains its own computing infrastructure in the form of the \ldg
that supplied the majority of the tens of millions of core hours of
computing time used in the two previous observation campaigns. 
The \ldg provides a homogeneous compute environment for gravitational wave data analysis.
It consists of a single version of CentOS Linux as the operation system, and provides a
software stack mandated by LIGO. 

The observation of gravitational waves with an international network of
gravitational wave detectors was accomplished in LIGO-Virgo's
second observing run, known as O2. In the upcoming third observing run O3,
the Japanese KAGRA detector will further expand this network. Ongoing improvements
to the sensitivity of these observatories, coupled with longer discovery campaigns,
will exacerbate the need for computational resources, in particular for 
low-latency (order of seconds to minutes) searches.  
Anticipating this scenario, LIGO has expanded its \ldg to exploit
additional resources. In the following section we describe recent 
deliverables of this effort.

\subsection{High-energy Particle Physics}
\label{sec:hep}

The goal of particle physics is to understand the universe at its most fundamental level, including the constituents of matter and their interactions. Our best theory of nature---the standard model (SM)---is a quantum field theory (QFT) that describes the strong, electromagnetic (EM) and weak interactions among fundamental particles, which are described as {\it fields}. In the SM, the weak and EM forces have the same strength at very high energy (as existed in the early Universe) described by single {\it electroweak} interaction and particles must be massless to preserve {\it gauge invariance} in which different configurations of the fields lead to identical physics results. Gauge invariance is a required ingredient of any QFT describing nature, otherwise calculated values of physically measurable quantities, such as the probability of particles scattering with one another at high energy, can be infinite. Since we know through observation that the EM force is much stronger than the weak force, electroweak symmetry is a {\it broken symmetry}. We also know that most fundamental particles have mass, including the weak force carriers that are massive and have a short-ranged interaction. Exactly how this symmetry is broken and how fundamental particles acquire their mass without violating gauge invariance is one of the most important questions in particle physics.

The SM provides a mechanism that answers both of these questions simultaneously. Particle masses arise when the electroweak symmetry is spontaneously broken by the interaction of massless fields with the {\it Higgs field}, an invisible, spinless field that permeates all space and has a {\it non-zero value everywhere}, even in its lowest energy state. A would-be massless particle that interacts with the Higgs field is slowed down from the speed-of-light due to this interaction and consequently acquires a non-zero mass. This {\it Higgs mechanism} makes the remarkable prediction that a {\it single massive, neutral, spinless particle} called the ``Higgs boson''---a quantum excitation of the Higgs field---{\it must exist}~\cite{PhysRevLett.13.321,HIGGS1964132,PhysRevLett.13.508,PhysRevLett.13.585,PhysRev.145.1156,PhysRev.155.1554}.

Equally remarkable is that we have progressed technologically to be able to produce Higgs bosons in the laboratory. While Higgs boson mass is not predicted by the SM, it must be less than $\sim$1000 times the mass of the proton to avoid the infinities previously mentioned. In Einstein's special relativity, a mass \(m\) is equivalent to an energy content \(E\) through the relation \(E=m\,c^2\)~\cite{1952E}. Particle accelerators can impart energy to particles to form an equivalent amount of mass when they are collided. The Large Hadron Collider (LHC) at CERN in Geneva, Switzerland is the world's most powerful particle collider and was built with the primary goal of either discovering the Higgs boson or refuting its existence. At the LHC, counter-rotating bunches of 10$^{11}$ protons are accelerated inside a 27-km circular ring and focused to collide at rate of 40 MHz with a (design) center-of-mass energy of 14 trillion electron-volts. This is the energy equivalent of the rest mass of 14,000 protons, which is sufficient to excite the Higgs field to produce Higgs bosons~\cite{higgs1,higgs2,higgs3,higgs4}. Higgs bosons decay almost immediately and sophisticated detectors surrounding the collision region are used to detect and measure their decay products, enable physicists to piece them together to search for Higgs boson production within the LHC.

In 2012, the ATLAS and CMS collaborations announced the discovery of a Higgs boson at the LHC~\cite{HIGG-2012-27,Chatrchyan:2012xdj}. This discovery lead to the 2013 Nobel Prize in Physics for the theory of the Higgs mechanism and prediction of the Higgs boson. Since this discovery, the properties of this new particle---its mass, spin, couplings to other SM particles, and certain symmetry properties---have been measured with increasing precision and found to agree with the SM prediction.

The development of the SM is a triumph of 20$^{\rm th}$ century physics, with the last piece of the puzzle put in place by the discovery of the Higgs boson. \co{However, this is far} from the end of the story of particle physics. Even without including gravity, we know that the SM is an incomplete description of nature and leaves many open questions to be answered. For example, the SM does not include non-zero neutrino mass and mixing which is observed in solar and atmospheric neutrino experiments~\cite{Ahmad:2001an,Fukuda:1998mi}, nor does it account for the predominance of matter over antimatter. Moreover, the SM accounts for only 5\% of the known mass-energy content of the universe and does not describe dark matter or dark energy that comprises the rest.

The LHC will continue to provide a unique window into the subatomic world to pursue answers to these questions and study processes that took place only a tiny instant of time after the Big Bang. The next phase of this global scientific endeavor will be the High-Luminosity LHC (HL-LHC) which will collect data starting circa 2026 and continue into the 2030s. The goal is to search for physics beyond the SM and, should it be discovered, to study its details and implications. During the HL-LHC era, the ATLAS and CMS experiments will record $\sim$10 times as much data from $\sim$100 times as many collisions as was used to discover the Higgs boson, raising the prospect of exciting discoveries during the HL-LHC era.

\subsubsection*{ATLAS Data Analysis with HTC workloads on HTC infrastructures}
\label{sec:hep-atlas-HTC}

The ATLAS detector~\cite{ATLASDetector} is a multi-purpose particle detector at the LHC with a forward-backward symmetric cylindrical geometry and a nearly $4\pi$ solid angle coverage of the LHC collision region. 
The ATLAS detector is eight stories tall, weighs 7000 tonnes and consists of $\sim$100 million electronics channels.
At a proton bunch crossing rate of 40 MHz, there are $\sim$1 billion proton-proton interactions per second occurring within the ATLAS detector. The rate of data generated by the detector is far too high to collect all of these collisions, so a sophisticated trigger system is employed to decide which events are sufficiently interesting for offline analysis. On average, only 1 in every 100,000 collisions is archived for offline analysis. A first-level trigger is implemented in hardware and uses a subset of the detector information to reduce the accepted rate to a peak value of 70~kHz. This is followed by a software-based trigger run on a computing cluster that reduces the average recorded collision rate to 1~kHz.

With $\sim$$50\,\mathrm{PB}$ of data generated annually by the LHC experiments, processing, analyzing, and sharing the data with thousands of physicists around the world is an enormous challenge. To translate the observed data into insights about fundamental physics, the important quantum mechanical processes and response of the detector to them need to be simulated to a high-level of detail and with a high-degree of accuracy.

Historically, the ATLAS experiment has used a geographically distributed grid of approximately 200,000 cores continuously (250,000 cores at peak) to process, simulate, and analyze its data. The ATLAS experiment is currently responsible for 1,000 million core-hours per year for processing, simulation, and analysis of data, with more than 300 PB of active data.
In spite of these capabilities, the unprecedented needs of ATLAS have led to contention for computing resources. The shortfall became particularly acute in 2016-17, as the LHC delivered about 50\% more data than planned, and the LHC continues to generate more data than planned. The shortfall will not be met by growth from Moore's Law, or simply more dollars to buy resources. Furthermore, once the HL-LHC starts producing data in 2020 this problem will only be magnified, making the gap more acute.

A partial response to these challenges has been to move from utilizing infrastructure that was exclusively distributed and that only supported the HTC mode, to an infrastructure
mix that also included HPC infrastructure such as the Blue Waters and Titan supercomputers. 

\co{These leadership class HPC facilities are geared towards supporting a
workload mix consisting of relatively few concurrently running jobs, each of
which uses a significant fraction of the system resources. Combined with the
requirement to choose nodes that are close to each other in the HPC network when
selecting nodes to run a job, increases scheduling difficulty and reduces the
overall utilization of the cluster. In the case of Blue Waters implementing
this \emph{topology aware scheduling} reducing cluster utilization to $\sim
80\%$ while \emph{increasing} science output due to the increase in simulation
speed~\cite{jenos:2016}.}

This expansion of infrastructure types has been primarily motivated by 
the very practical need to alleviate the ``resource scarcity'' as the requirements of ATLAS have
continued to grow. This has required addressing both intellectual and technical challenges of using HPC infrastructure in a HTC mode. Section~\ref{sec:hep-atlas-prodBW} discusses these challenges and the ATLAS project's response to them.

\section{Open Science Grid and containers pave the way to create a unified HTC and HPC infrastructure} 
\label{htc_on_hpc}

\texttt{OSG} provides federated access to
compute resources for data-intensive research across science
domains, and it is primarily used in physics. Workloads that best use this large pool of resources need to meet clearly defined criteria:  
\begin{itemize}
\item They consist of loosely coupled jobs that require a few cores to
at most one node.
\item Furthermore, since compute resources are 
not owned by \texttt{OSG}, and jobs may be killed and re-started 
at different sites when higher priority jobs enter the system, workflow managers that can preempt a job without
losing the work the job has already accomplished should be used.
\item Additionally,
jobs should be single-threaded and require less than 2~GB of memory
in each invocation, and which can run for up to twelve hours.
\item Input and output
data for each job is limited to
10~GB.
\end{itemize}

\noindent Another important consideration is that \texttt{OSG} resources do not typically 
have the same software ecosystem required by LIGO or ATLAS workloads. 
This has led to the development of software stacks that seamlessly run on
disparate compute resources. For desktop and server applications, 
Docker containers have become one of the preferred solutions to address this problem
of encapsulating all required software dependencies of an application and
providing a uniform way to share these packages. 

The high energy physics community has made extensive use of containers 
to run ATLAS workloads on disparate \texttt{OSG} compute resources. \texttt{Shifter}, and more
recently Singularity, have been implemented as container solutions by the \texttt{OSG} project.
Similarly, LIGO scientists containerized their most compute-intensive HTC workload to 
seamlessly run on \texttt{OSG} resources. As a result, \texttt{OSG} contributed about 10\% of the compute time consumed 
during LIGO-Virgo's first and second discovery campaigns. Both \texttt{Shifter} and Singularity 
are currently used to containerized LIGO's software stacks.

ATLAS and LIGO scientists soon realized that the approach used to
connect their HTC infrastructure to the \texttt{OSG} could also be used to construct a unified HTC-HPC infrastructure. 
To accomplish this, containers were deployed on HPC infrastructures, which were then configured as \texttt{OSG} compute elements. This \co{approach} enabled a seamless use of HPC infrastructures for ATLAS and LIGO large scale data analyses. In the following section we describe how these two milestones were accomplished, highlighting the similarities and differences between these approaches.

\subsection{Scaling gravitational wave discovery with \texttt{OSG} and \texttt{Shifter} connecting the \ldg to Blue Waters}

In the last two LIGO and Virgo observing runs, the LDG
benefited from adopting \texttt{OSG} as a universal adapter to external resources,
increasing its pool of compute resources to include campus and regional clusters,
the NSF funded Extreme Science and Engineering 
Discovery Environment (XSEDE)~\cite{2017Weitzel}, and opportunistic cycles from US Department of
Energy Laboratories and High Energy Physics clusters. 

\co{In order} to connect the \ldg to Blue Waters, the NSF-supported, leadership-class
supercomputer operated by NCSA,
authors of this article spearheaded the unification of \texttt{OSG}, \texttt{Shifter}, and Blue Waters~\cite{BOSS:2017,shifter}.

\begin{figure}[htbp]
\centering
\includegraphics[height=0.6\linewidth]{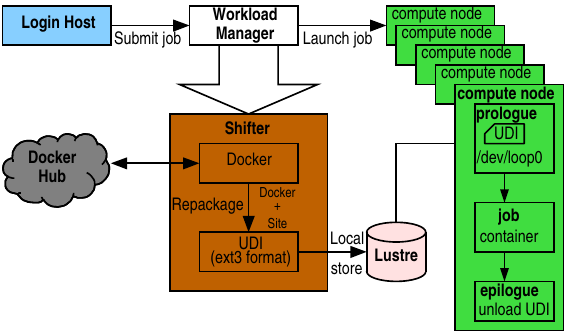}
\caption{The components involved in starting a \texttt{Shifter} job on Blue Waters.
Jobs are submitted to the workload manager on Blue Waters' login nodes, which
launches jobs on the compute nodes. For jobs requesting the use of containers,
the workload manager first instructs the \texttt{Shifter} runtime environment to pull
an up-to-date copy of the container image from Docker Hub. The container image
is repackaged as a user defined image using a regular squashfs disk image. Finally,
the disk image is loop-mounted by the jobs on the compute nodes during their
prologue and unloaded in the epilogue after the job ends.}
\label{fig:shifter-ligo}
\end{figure}

Since \texttt{Shifter} is supported natively by Blue Waters, LIGO used it
to encapsulate a full analysis software stack and to use Blue Waters as a
computing resource during O2~\cite{BOSS:2017}, 
 adding gravitational wave science to the
portfolio of science enabled by Blue Waters. 
Figure~\ref{fig:shifter-ligo} shows the setup
used to start \texttt{Shifter} jobs.

To further leverage existing efforts to use cycles on HPC clusters for HTC
workloads, LIGO decided to base its efforts on the existing \texttt{OSG} infrastructure
and to create a container that can be used on any \texttt{OSG} resource along with
\co{methods to use} Blue Waters as an \texttt{OSG} resource provider.

\begin{figure}[htbp]
\centering
\includegraphics[width=0.9\linewidth]{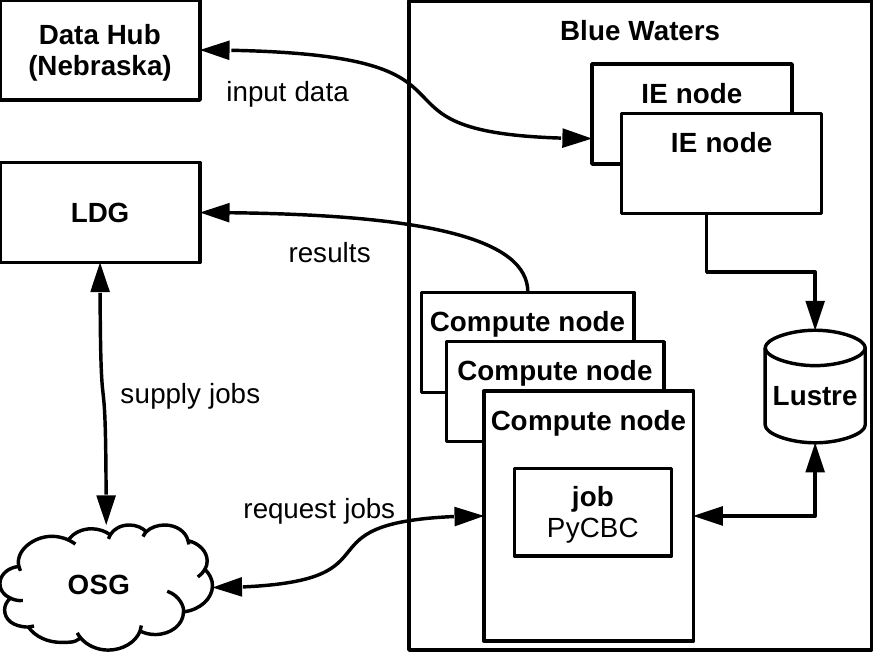}
\caption{Interaction between the LIGO Data Grid, the Open Science Grid, and
\texttt{PyCBC} jobs during a detection run. Pilot jobs are started on Blue Waters
compute nodes, which register themselves with \texttt{OSG} and request compute jobs. 
The LDG-hosted Condor submission host supplies compute jobs to \texttt{OSG} to be executed by the Blue
Waters workers. Once a worker starts up, it requests data to be processed from
the data hub hosted at Nebraska Supercomputer center and returns results to
LIGO.}\label{fig:osg-interaction}
\end{figure}

The solution we have developed to use Blue Waters, or any other HPC infrastructure, is as
follows (also see Figure~\ref{fig:osg-interaction})~\cite{BOSS:2017}:
\begin{itemize}
\item LIGO data analysis jobs are submitted to the \htcondor scheduler running
at an \ldg site, which oversees the workload and schedules work items on either
the local compute resources or on remote resources.
\item \glidein~\cite{sfiligoi2009pilot} pilots are submitted as regular jobs
to the HPC cluster's job
scheduler to reserve a number of compute nodes for use by \texttt{OSG}. This creates a
virtual private batch systems that reports back to the \glidein
Workload Management System at the \ldg site to temporarily become part of the
\htcondor worker pool. 
\item \co{Once the most compute-intensive jobs in the \texttt{OSG} pool start flowing into Blue Waters,  
LIGO data is transferred at scale from the data hub hosted at the Nebraska supercomputer 
center into Blue Waters. This is 
accomplished by using a distributed data access infrastructure, 
which is based on the \texttt{XRootD} server suite and \cvm~\cite{2017Weitzel}. In practice, Nebraska's 
data transfer node (DTN) endpoint is used to distribute the data. Each of these 12 DTNs has a single 
10Gbps interface, and utilizes the Linux Virtual Server~\cite{lvs} to provide a single, load-balanced IP address. 
Furthermore, given that LIGO data is restricted to LIGO members, it is necessary to use the secure \texttt{CVMFS} mode, i.e., X509 certificates, to authenticate and authorize users to access the data.}
\item \co{Once the data has been transferred,} jobs are run on the \glidein workers until all have completed or
the pilot jobs expire. \co{Thereafter, and as part of the compute workflow, the data products are transferred back to the host LIGO cluster from which the workload was launched. It is worth highlighting that, at present, this approach is only amenable to Pegasus~\cite{deelman2005pegasus} workflows. However, there are ongoing efforts to use Rucio~\cite{rucio} for LIGO 
bulk data management~\cite{ligorucio}.}
\end{itemize}

\begin{figure*}[!ht]
\centerline{
  \includegraphics[width=0.42\textwidth]{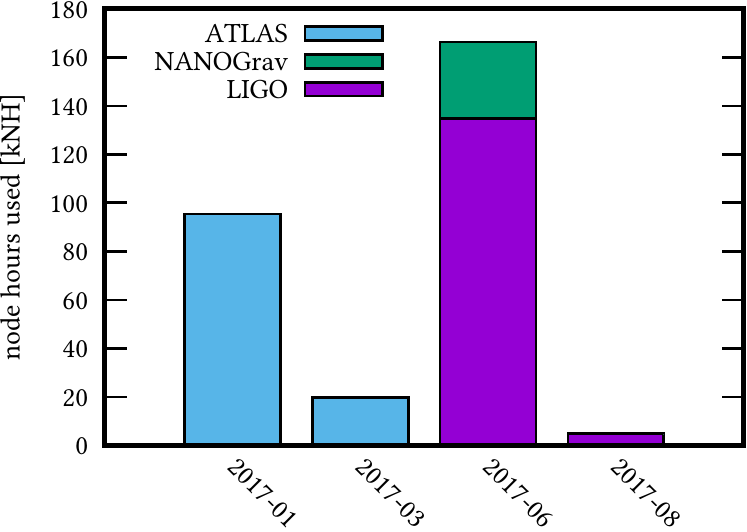}
 \raisebox{6mm}{%
  \includegraphics[height=0.27\textwidth]{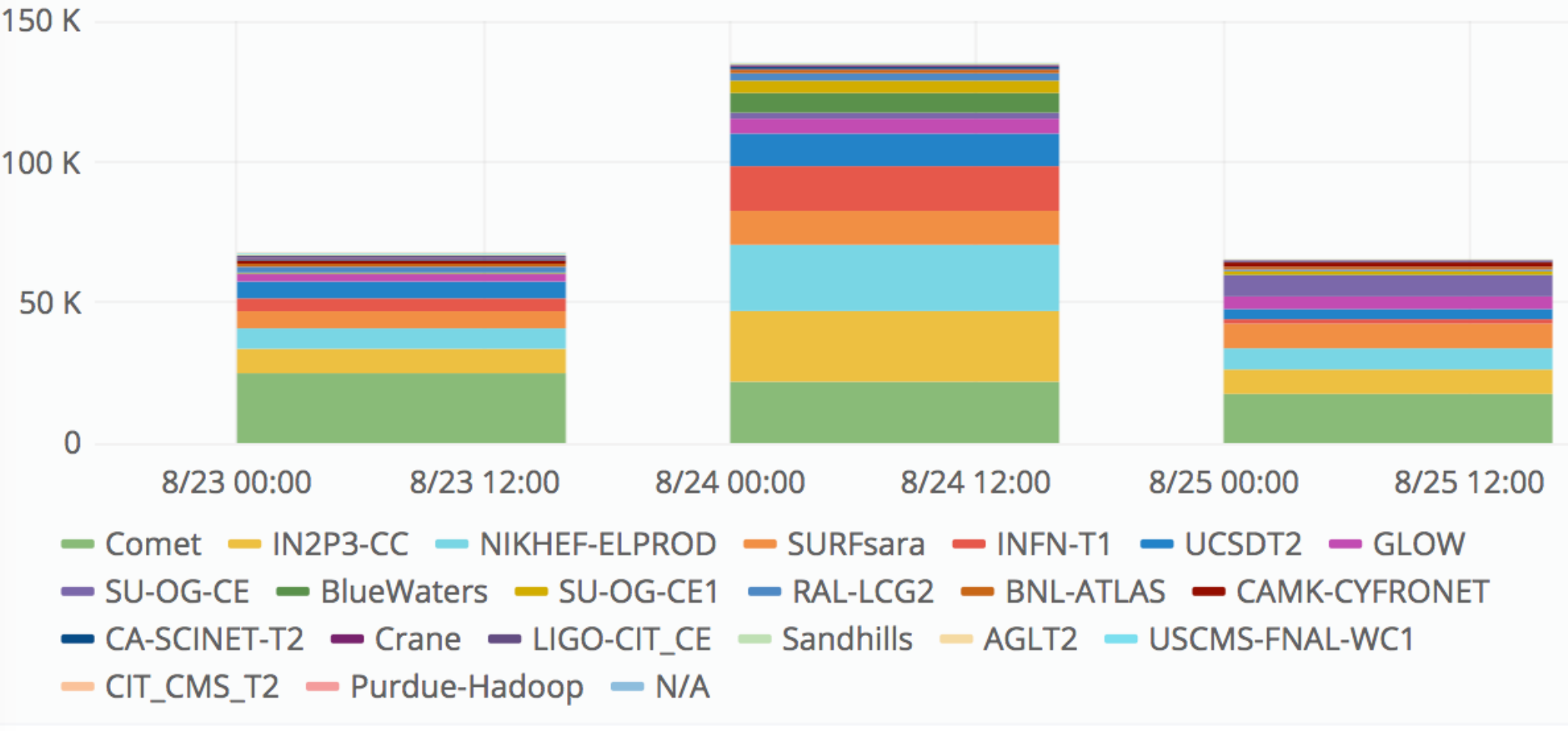}}
  }
\caption{Left panel: gravitational wave astrophysics (LIGO and NANOGrav) and high energy physics (ATLAS) projects that use HPC containers to seamlessly exploit the unique computing capabilities of the Blue Waters supercomputer. Right panel: Open Science Grid compute resources used for large-scale gravitational wave data analysis. The chart shows the first time Blue Waters was used at scale as an Open Science Grid compute element, which corresponds to the gravitational wave discovery of two colliding neutron stars by the LIGO and Virgo detectors.}
\label{fig:2}       % Give a unique label
\end{figure*}

\noindent As shown in Figure~\ref{fig:2}, the first time this framework was used
for a production scale analysis \co{on Blue Waters} was for the validation of the first 
gravitational detection of two colliding neutron
stars by the LIGO and Virgo detectors, an event that marked the
beginning of multimessenger astronomy~\cite{bnsdet:2017,mma:2017arXiv,grb:2017ApJ}. 
\co{The open source, \texttt{PyCBC} pipeline used for this work can be obtained at~\cite{pycbccite}. 
Documentation to use this pipeline on \texttt{OSG} resources is available at~\cite{pycbcosg}, whereas instructions to 
run \texttt{PyCBC} on Blue Waters using \texttt{Shifter} and \texttt{OSG} is presented in~\cite{BOSS:2017,shifter}. Recent developments to provide access to XSEDE resources through \texttt{OSG} infrastructure is described in~\cite{Thapa:2018}. 
Detailed information to launch these workloads can be found at~\cite{edpres}.}

This work has multifold implications. First of all, it provides an additional pool of 
computational resources that has already been used to promptly validate major scientific discoveries. 
In the near future, Blue Waters will continue to provide resources to accelerate gravitational wave searches, 
and to enable computational analyses beyond the core investigations that may lead to new insights through
follow up analyses. This work will also benefit cluster utilization without affecting the network performance of HPC jobs. 
More importantly, this success clearly exhibits the interoperability of NSF cyberinfrastructure resources, and makes 
a significant step to further the goals of the US National Strategic Computing Initiative, i.e., to foster the convergence 
of data analytic computing, modeling and simulation. Supporting high throughput LIGO data analysis workloads 
concurrently with highly parallel numerical relativity simulations and many other complex workloads is the most 
recent success and most complex example of successfully achieving convergence on Leadership Class computers 
like Blue Waters, which is much earlier than was expected to be possible.

\subsection{\texttt{OSG} and containers in Blue Waters for High Energy Physics}
\label{sec:hep-atlas-prodBW}

\begin{figure}[htbp]
\centering
\includegraphics[height=0.8\linewidth]{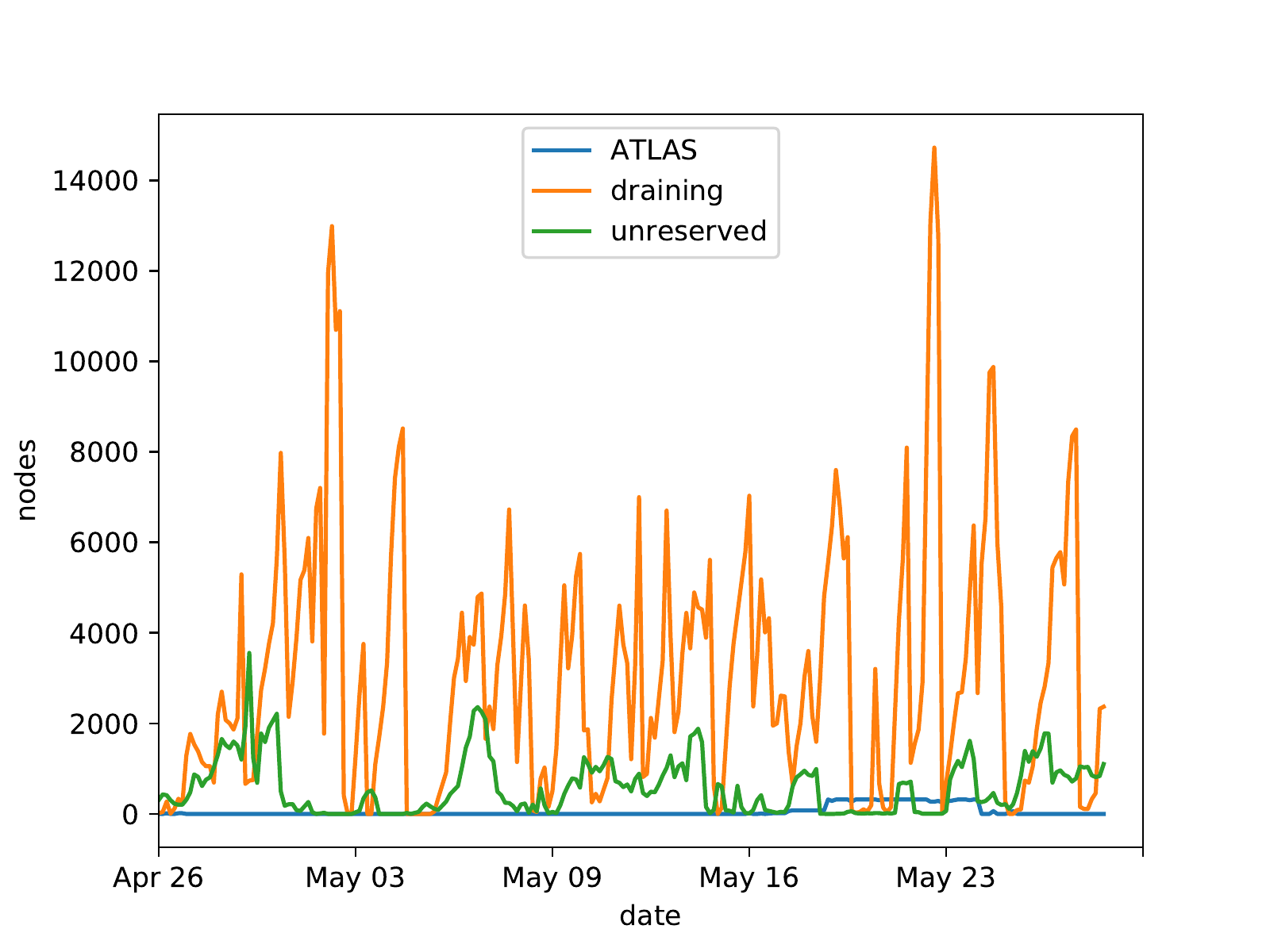}
\includegraphics[width=\linewidth]{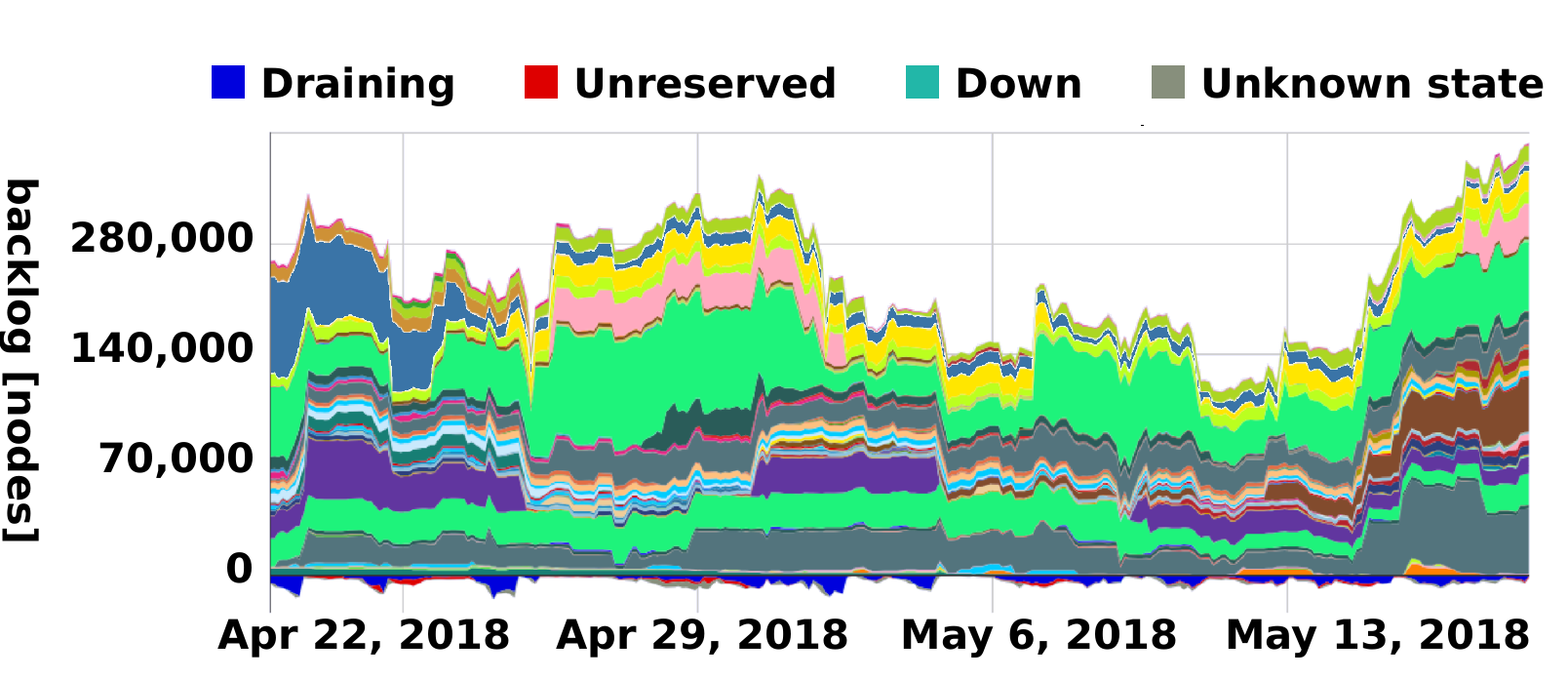}
\caption{\co{Top panel: period of time during which 
35 million ATLAS events were processed using 300 Blue
Waters nodes. Utilization during this period averaged to be $81\%$, which is typical
for Blue Waters. Bottom panel: backlog of queued up jobs for the same period
in requested nodes, with colors indicating user accounts. During this period
the queued up workload never dropped below 80,000 nodes i.e., 4 times the
number of nodes in Blue Waters. The red and blue curve below the horizontal
axis are nodes available for work scavenging during this period. Plots taken
from~\cite{bauer:2018}.}}
\label{fig:BWUsageATLASPeriod}
\end{figure}

\noindent To simulate and process large amount of data from the LHC, Blue Waters has been integrated into the ATLAS production processing environment by leveraging \texttt{OSG CONNECT} and \texttt{MWT2} services. ATLAS jobs require a specific environment on the target site to execute properly. These include a variant of the CentOS6 operating system, numerous RPM packages and the distribution of ATLAS software libraries via \texttt{CVMFS} repositories. Blue Waters compute nodes themselves do not provide the required environment for ATLAS jobs as they use an older SUSE OS variant nor do they include many of needed RPM packages. Docker Images are delivered via \texttt{Shifter} to create an environment on Blue Waters nodes that are compatible with the ATLAS job payload. Though \texttt{CVMFS} cannot be used directly due to a lack of FUSE availability on Blue Waters, access to on-disk copies of the repositories is made available via a softlink from the required root of \texttt{CVMFS} to the location of the local repositories created by an rsync-based \texttt{CVMFS} replication service. To comply with Blue Waters' two factor authentication, the RSA One Time Password (OTP) authentication system is used to create a proxy valid for 11 days. The OTP-based proxy is renewed on a weekly basis using MyProxy~\cite{MyProxy}. The OpenSSH client (gsissh) uses this proxy to ssh into a Blue Waters login node and startup SSH glideins for an HTCondor overlay that is used to schedule ATLAS jobs on Blue Waters. 

The ATLAS jobs start flowing into Blue Waters when glideins submitted within a container in Blue Waters contact a Production and Distributed Analysis 
 (PanDA) workload management system~\cite{Maeno:2011zz} at CERN to get an ATLAS payload. The glideins also pull all the necessary data and files using the Local Site Mover (LSM) at the University of Chicago. To minimize network transfer on stage-in, data is cached to the Blue Waters local Lustre file system.  When the ATLAS jobs run, they use the stage-in data for their input and write their output back to the Lustre scratch disk. Once the workload is complete, the output data is transferred to the data storage system at the University of Chicago. \co{Further information and documentation about the virtual cluster system utilized for this approach is described in~\cite{atlasconnect,jaya:2015J}.}

\co{Figure~\ref{fig:BWUsageATLASPeriod} shows a particular one month period in 2018 in which 35k Blue Waters cores were utilized (peak) to process 35M collision events. The top panel of this figure shows that this approach is cost-effective, boosting cluster utilization, and has no adverse effect on other HPC workloads.} The job output was made available to the rest of the ATLAS collaboration for use in analysis of the LHC data to improve SM measurements and to search for new physics beyond the SM.

\subsection{ATLAS \& PanDA \& Titan}
\label{sec:hep-atlas-usage}

The computing systems used by LHC experiments has historically consisted of the federation of hundreds to thousands of distributed resources, ranging from small to mid-size resource. In spite of the impressive scale of the existing distributed computing solutions, the federation of small to mid-size resources has proven to be insufficient to meet current and projected future demands. 

\begin{figure}[htbp]
\centering
\includegraphics[width=\linewidth]{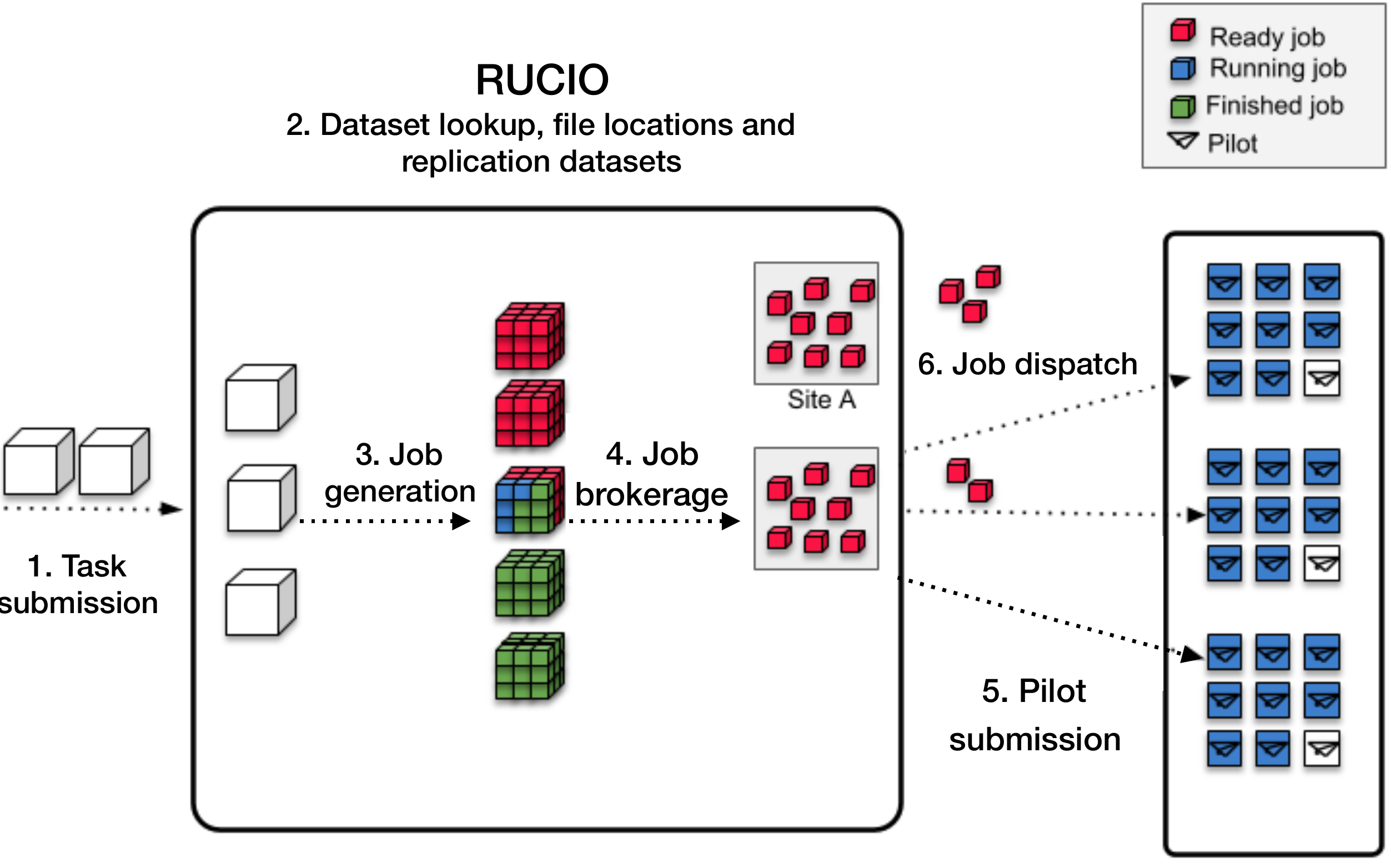}\hspace{-4em}%
\caption{Schematic showing the primary stages in execution of ATLAS workloads on Titan using the BigPanDa workload management system. PanDA's broker acts on jobs (as opposed to tasks), and uses their description to determine how best to insert aggregate and shape into existing backfill slots on Titan. Although not used to submit to non-Oak Ridge Leadership Computing Facility sites, in principle the PanDA broker could set jobs to go to another resource (Site A).}\label{fig:panda-atlas-titan}
\end{figure}

The ATLAS experiment has embraced Titan, a US Department of Energy (DOE) leadership facility, in conjunction with traditional distributed high-throughput computing to reach sustained production scales approaching 100M core-hours a years (in 2017),
and easily surpassing 100M in 2018. 
Underpinning these efforts has been the PanDA workload management system, which was
extended to support the execution of ATLAS workloads on Titan~\cite{htchpc2017converging}, as shown in Figure~\ref{fig:panda-atlas-titan}.
This work initially critically evaluated the design and operational considerations needed to support the sustained, scalable and production usage of Titan for ATLAS workloads in a high-throughput mode using 
the ``backfill'' operational mode.   
It also preliminarily characterized a next generation executor for PanDA to support new workloads and advanced execution modes
as well as outlining early lessons for how current and future experimental and observational systems can be integrated with production supercomputers and other infrastructures in a general and extensible manner.

As shown in Figure~\ref{fig:panda-atlas-titan}, ATLAS payloads use Titan compute resources as follows: PanDA pilots run on Titan's DTNs. This is advantageous, since DTNs can communicate with the PanDa server through a fast internet connection (10-GB/s). Furthermore, the worker nodes on Titan and the DTNs use a shared file system, which allows the pilots to stage in data and files that are needed by the payload, and to stage out data products once the payload is completed. PanDA pilots query Titan's Moab scheduler to check whether available resources are suitable for PanDA jobs, and transfers this information to the PanDA server which then prepares a list of jobs that can be submitted on Titan. Thereafter, the pilot transfers all the necessary input data from Brookhaven National Laboratory, a Tier 1 ATLAS computer center.  \co{Additional documentation of this approach, including progress reports, can be found at~\cite{htchpc2017converging,reportkatz}. Detailed documentation on the the PanDA Production and Distributed Analysis System is available at~\cite{Maeno:2011zz,pandadoc,maeno}.}

\section{Analysis of the case studies}
\label{sec:analysis}

In this section we discuss similarities and differences between the case studies. We start by identifying similarities in the approaches followed by the high energy physics and gravitational wave communities to run HTC-type workloads in the Blue Waters supercomputer 

\subsection{Similarities between case studies}

We have identified the following common features between LIGO and ATLAS workloads that utilize Blue Waters

\begin{enumerate}
\item \texttt{CVMFS} and/or \texttt{XRootD} is used for global distribution of software and data
\item \texttt{Shifter} is used as a container solution for both software stacks
\item ATLAS and LIGO workloads are planned targeting Blue Waters as an \texttt{OSG} compute element
\item Jobs submitted to the \texttt{OSG} will start flowing into Blue Waters when glideins are started within a \texttt{Shifter} container in Blue Waters
\item These workloads use HTCondor to schedule jobs. LIGO workloads also use Pegasus~\cite{deelman2005pegasus} as a workflow management system
\item The  workloads use temporary certificates to comply with two factor authentication
\item The \texttt{OSG} is used as a global adapter to connect ATLAS and LIGO compute-resources to Blue Waters
\item These workloads use the backfill operational mode to maximize cluster utilization without loss of overall quality-of-service
\end{enumerate}

\subsection{Differences between case studies}

In this section we focus on the ATLAS workload designed to run at production scale in the Titan supercomputer. The differences between this LHC workload and those discussed in the previous section are:

\begin{enumerate}
\item Instead of HTCondor, this workload uses PanDA as the workload management system
\item It targets Titan, a US DOE leadership-class supercomputer, to reach sustained production of 51M core-hours per year
\item PanDA brokers were deployed on Titan to enable distributed computing at scale
\item PanDA Broker pulls jobs' input files from Brookhaven National Laboratory Data Center to the Oak Ridge Leadership Computing Facility (OLCF) Lustre file system. On the other hand, LIGO and ATLAS workloads that utilize Blue Waters, transfer data at scale from Nebraska and the University of Chicago, respectively
\item PanDA Brokers are deployed on DTNs because these nodes are part of the OLCF infrastructure and can access Titan without RSA SecureID authentication. DTNs are not part of Titan's worker nodes and, therefore, are not used to execute Titan's jobs
\item PanDA Broker queries Titan's Moab scheduler about the current available backfill slot, and creates an MPI script, wrapping enough ATLAS jobs' payload to fit the backfill slot. Thereafter, PanDA Broker submits the MPI script to the Titan's PBS batch system as shown in Figure~\ref{fig:panda-atlas-titan}. In contrast, ATLAS and LIGO workloads in Blue Waters use the
\texttt{COMMTRANSPARENT} flag, so that each task can be placed anywhere within   the   torus   network   without   affecting   the   network performance of other jobs, and increasing the overall system utilization
\item Once every MPI script is finished, PanDA Broker transfers the data products to Brookhaven National Laboratory. In contrast, LIGO data products on Blue Waters are transferred back to the host \ldg cluster, and ATLAS workloads using Blue Waters resources stageout data products to the data storage space at the University of Chicago
\end{enumerate}

\section{Exascale Computing: Scope and future applications}
\label{sec:5}

In early 2019, the LIGO, Virgo and KAGRA detectors will 
gather data concurrently for the first time. This one-year campaign will
benefit from ongoing commissioning work at the LIGO and Virgo sites. The implications of
this are multifold. First of all, more sensitive detectors means that gravitational wave
signals will spend more time in the detectors' sensitive frequency range. In turn, effectual 
searches will require many more waveforms that are significantly longer than
in previous campaigns. 

Additionally, more sensitive detectors means that they can probe
a larger volume of the Universe, which will boost the number of sources that will be detected.
From a data analysis perspective, this means that
we will require a significant increase in the pool of computational resources to keep the same
cycle of detection to publication. If the detection rate increases by at least 
a factor of two, this 
level of activity will become unsustainable. Requiring that new data becomes publicly available
with a six month latency also implies that compute-power will be utilized to address core data 
analysis activities, at the expense of not pursuing high risk-high reward science investigations
that may lead to groundbreaking discoveries.

This situation is not unique to gravitational wave data analysis. For the HEP, the data volumes to be processed by 2022 (Run 3) and then the HL-HLC starts producing data (Run 4) will increase by factor of 10-100 compared to the existing volumes (Run 2).  It is acknowledged in the HPC community 
that there is a growing disconnect between commercial clouds and HPC infrastructures, where the computing power and
data storage concentrate, and edge environments which are experiencing the largest increase of data volumes but lack the needed infrastructure to cope with it. In this scenario, LIGO and ATLAS represent an edge environment that will generate very large datasets in the very near future, and will require access to ever increasing pools of computational power. 

\co{Another large-scale facility that will rapidly become a top user of computing resources is the LSST, which is expected to start operations in the early part of the next decade. This survey will produce a stream of \(\sim10\) million time-domain events per night, which will be transmitted to NCSA, using a dedicated network, within 60 seconds of observation. These observations will encompass nearly 6 million bodies in the Solar System, 20 billion galaxies, 17 billion stars, 7 trillion observations, and 30 trillion measurements produced annually~\cite{lsstbook}. Since the primary objectives of the LSST project are to acquire, process and make available the stream of event alerts and data release data products to data-rights holders, it is planned that the LSST project will provide a portal to explore, subset and visualize the LSST Archive. A small cluster with approximately 2,400 cores, 18TFLOPS, 4PB of file storage and 3PB of database storage will be provided for these light-weight data analyses. It is worth mentioning that these resources may support up to one hundred users accessing the cluster concurrently. Users who need larger resources will need to obtain computing resources elsewhere.}

\co{To put in perspective the amount of computing power that will be needed for LSST science, we can take as a reference point the ongoing Dark Energy Survey (DES)~\cite{DES:2016MNRAS}, which is regarded as the precursor of LSST. The DES Data Processing and Calibration System, which is used for automated processing and quality control across multiple images covering a footprint of 5,000 \(\textrm{deg}^2\). These processes are currently carried out on dedicated development clusters and HPC platforms across the US and Germany~\cite{Mohr:2012SPIE}, which include Blue Waters~\cite{Kramer2015,bluewaters:web}, the Illinois Campus Cluster Program~\cite{CCUUIUC} and the \texttt{OSG} at Fermilab~\cite{jaya:2015J}. Note that these computational resources are only used for image processing, which is often followed up by in-depth statistical analyses, and large-scale simulations to compare existing models of the structure and evolution of the Universe with actual DES data~\cite{Sunayama:2016JCAP,Li:2016ApJ,DESI:2016D}. Having established this baseline of minimal computational resources needed for core DES data analyses, one can now place this in perspective by considering that LSST will cover 20,000 \(\textrm{deg}^2\), with an improved distance of about 2.5. In brief, the large-data volumes produced at such high cadence will require a significantly larger pool of computing resources for low-latency studies of transient sources. In addition to these requirements, we need to consider that large-scale simulations will be needed to compare the predictions of existing cosmological models~\cite{Lawrence:2017ApJL,Emberson:2018E} with LSST's ultra high-definition observations.}

A range of opportunities have been discussed in the HPC community to alleviate these challenges. Some of them include data processing as close as possible to the data sources, and logically centered cloud-like processing. The use of containers will continue to play a significant role to seamlessly run compute-intensive workloads on commercial clouds, HPC infrastructures, and computing resources deployed in edge environments where the datasets are generated. The development of a common interface for containerization will facilitate convergence for all the ecosystem of applications that scientific cyberinfrastructure has to address. \co{It is expected that future HPC platforms will provide the necessary flexibility to run HTC workloads through backfilling. The Frontera supercomputer~\cite{Front} already envisions these type of activities in conjunction with XSEDE systems~\cite{xsede,ecss}.}

\subsection*{Outlook}

As the \co{data revolution} continues to evolve, new paradigms will emerge to support compute-intensive and data-intensive work either in HPC centers or edge environments. Global recommendations from the HPC communities for edge environments include the development of new algorithms to compress datasets by one or more orders of magnitude, and to understand how to use lossy compression. Furthermore, next-generation workloads may include not only classical HPC-type applications, but also machine and deep learning applications, which require a new level of abstraction between software and hardware to run these type of hybrid workloads. As HPC and the big data revolution continue to develop and converge, new needs and opportunities will arise, including the use of HPC math libraries for high end data analysis, the development of new standards for shared memory, and the interoperability between programming models and data formats.  

The data revolution has already initiated a paradigm shift in gravitational wave astrophysics and high-energy physics. Deep learning algorithms have been used to show that gravitational wave detection can be carried out faster than real-time, while also increasing the depth and speed of established LIGO detection algorithms, and enabling the detection of new classes of gravitational wave sources~\cite{geodf:2017a,geodf:2017b,rebei:2018,hshen:2017,georgenoise,hall}. Deep learning approaches to the search for new physics at the LHC started around 2012 and has since been applied to address many challenges including simulation, particle identification, and event characterization~\cite{Guest:2018yhq}. These algorithms have been developed by combining HPC, innovative hardware architectures, and deep learning algorithms. The potential of this new wave of innovation as an alternative paradigm to combining HTC and HPC to cope with the ever increasing demand for computational infrastructure of edge environments will be discussed in future work.

\begin{acknowledgements}
This research is part of the Blue Waters sustained-petascale computing project, which is supported by the National Science Foundation (awards OCI-0725070 and ACI-1238993) and the State of Illinois. Blue Waters is a joint effort of the University of Illinois at Urbana-Champaign and its National Center for Supercomputing Applications. We thank Brett Bode, Greg Bauer, Jeremy Enos, HonWai Leong and William Kramer for useful interactions. On behalf of all authors, the corresponding author states that there is no conflict of interest.
\end{acknowledgements}

% BibTeX users please use one of
%\bibliographystyle{spbasic}      % basic style, author-year citations
%\bibliographystyle{spmpsci}      % mathematics and physical sciences
%\bibliographystyle{plain}
\bibliographystyle{spphys}       % APS-like style for physics
\bibliography{references}   % name your BibTeX data base

\end{document}